\documentstyle[aps,preprint]{revtex}

\begin{document}
\draft
\title{
\bf Off-diagonal Interactions, Hund's Rules and Pair-binding \\
in Hubbard Molecules
}
\author{
S. L. Sondhi, M. P. Gelfand\cite{mpg}, H. Q. Lin\cite{hql} and D. K. Campbell
}

\address{
Department of Physics \\
University of Illinois at Urbana-Champaign \\
Urbana, IL 61801-3080 \\
U.S.A.
}
\maketitle

\begin{abstract}
We have studied the effect of including nearest-neighbor, electron-electron
interactions, in
particular the off-diagonal (non density-density) terms, on the
spectra of truncated tetrahedral and icosahedral ``Hubbard molecules,''
focusing
on the relevance of these systems to the physics of doped C$_{60}$. Our
perturbation
theoretic and exact diagonalization results agree with previous work in
that the density-density term suppresses pair-binding. However, we find that
for the parameter values of interest for $C_{60}$ the off-diagonal terms {\em
enhance}
pair-binding, though not enough to offset the suppression due to
the density-density term. We also
find that the critical interaction strengths for the Hund's rules violating
level crossings in C$_{60}^{-2}$, C$_{60}^{-3}$ and C$_{60}^{-4}$ are
quite insensitive to the inclusion of these additional interactions.

\end{abstract}



\section{Introduction}

To account for the superconductivity of the
alkali fullerides $A_3$C$_{60}$ ($A$=K, Rb, Cs) Chakravarty, Gelfand and
Kivelson
(CGK) \cite{cgk} suggested a mechanism involving purely electronic (e-e)
interactions.
They showed how this might come about by studying  a simplified (Hubbard)
model of a C$_{60}$ molecule and arguing
that for a range of the atomic parameters the molecule
exhibits ``pair-binding'' \cite{fn1}: that is, a pair of
uncoupled monoanions, C$_{60}^-$ + C$_{60}^-$, is unstable with respect to
charge disproportionation into C$_{60}$ + C$_{60}^{-2}$ (and likewise for
trianions and quintanions, {\it mutatis mutandis}).
Weak electronic overlaps between molecules would then lead to
superconductivity.
The pair-binding phenomenon was apparently related to an unusual feature
of the low-lying spectra of C$_{60}$ di-, tri- and, and quadr-anions (all of
which have three icosahedral multiplets of states which are themselves
degenerate in the $U=0$ limit): namely,
that Hund's rule was violated and the ground states were those of
{\it minimum\/} spin and {\it minimum\/} orbital degeneracy.

The mechanism of superconductivity in the fullerides is currently not fully
understood \cite{mpg-rev}  and the relevance of the work of CGK to that
problem is controversial; we will not shed any light on that question directly.
Our interest lies in examining the robustness of the pair-binding
and violations of Hund's rules found
by CGK to e-e interactions not included in their model but
expected to exist in the real materials. This issue
is by no means straightforward since the quantities of interest are small
differences of large energies.
Our work was stimulated in part by work of Goff and
Phillips \cite{gp}, who studied the effect of
further-neighbor density-density interactions on pair-binding.

We begin by defining the extension of the Hubbard model studied in
this paper and the quantities of interest.
Next, we present the results of perturbative calculations for C$_{60}$.
We discuss the assignments of values for the various parameters and the
corresponding values for the pair-binding energy and level splittings.
To assess the validity of our approach we
follow the lead of White {\em et al.}~\cite{white} and present both
perturbative and exact diagonalization results for C$_{12}$, a mythical homolog
of C$_{60}$ that has a similar level structure at the fermi energy. We conclude
by summarizing our results and their implications for understanding the
properties of the fullerides.

\section{The Extended Hubbard Model}
We study the Hamiltonian
\begin{eqnarray}
H = &-t&\, \sum_{\langle ij \rangle} ( h_{ij} + h_{ji} )
+ U\, \sum_i n_{i \uparrow} n_{i \downarrow}
+ V\, \sum_{\langle ij \rangle} n_i n_j \nonumber \\
&+& W\, \sum_{\langle ij \rangle} (h_{ij} + h_{ji})^2
+ X\, \sum_{\langle ij \rangle} (h_{ij} + h_{ji})(n_i + n_j)
\label{hamil}
\end{eqnarray}
where $\langle ij \rangle$ runs over nearest-neighbor pairs,
$h_{ij}=\sum_\sigma c_{i\sigma}^\dagger c_{j\sigma}$,
$n_{i}=\sum_\sigma c_{i\sigma}^\dagger c_{i\sigma}$ and
the electronic orbitals are arranged either on the vertices of a
truncated tetrahedron (C$_{12}$) or on those of a truncated icosahedron
(C$_{60}$).
The interaction terms are, implicitly, normal ordered. Hence, appearances
notwithstanding, the $W$ and $X$ terms do {\em not} renormalize the hopping
and chemical potential at zeroth order.
For our qualitative purposes we ignore the difference in the
hopping and nearest neighbor terms between the
``short'' and ``long'' bonds, except that in the kinetic energy
they are not taken to be exactly equal (but rather, with a ratio of 1.001),
so as to split some accidental degeneracies and thus simplify the
computer programs somewhat \cite{fn2}.

The quantities of interest for us are the energies of the low-lying states and
the resulting set of pair-binding energies,
\begin{equation}
E_{\rm pair}^{(i)} = 2 E_i - E_{i-1} - E_{i+1} .
\end{equation}
Here $E_i$ is the ground state energy of the molecule with $i$ electrons in
excess of
charge neutrality; positive values of the $E_{\rm pair}$ indicate that charge
disproportionation is favored.
(The reader is warned that, in a slight abuse of this notation, we will also
use $E_i$/$E_{\rm pair}^{(i)}$
with further elaboration, such as ``$E_2$/$E_{\rm pair}^{(1)}$ for the $^1A_g$
state of C$_{60}^{-2}$,'' to indicate that energy/energy difference
even when the $^1A_g$ state is not the ground state.)

The model with $V=W=X=0$ was studied by CGK using second-order perturbation
theory
in $U/t$; they found that the $E_{\rm pair}^{(i)}$ with odd $i$ become positive
at
intermediate values of $U \ge 3.3$. The tendency of an added $V$ to suppress
pair-binding was noted by White {\em et al.}~\cite{white} and by Goff and
Phillips \cite{gp} who also included longer ranged density-density
interactions.

Our inclusion of the leading off-diagonal (non density-density) terms is
principally motivated by the observation that energy differences such
as $E_{\rm pair}^{(i)}$ and level splittings are
much more sensitive functions of the interaction parameters than the
energies of the molecular states involved in their definition, and hence one
may well need to keep track of interactions that nominally enter with
small coefficients \cite{fn3}. These terms arise when the
underlying density-density interaction is re-expressed in the Wannier basis
necessary to deriving the extended Hubbard model. The general matrix
element of the interaction ${\cal V}({\bf x})$ in the Wannier basis
$\phi_i({\bf x})$ is
\begin{equation}
\langle ij| {\cal V} |kl \rangle = \int d{\bf x}d{\bf y}\, \phi_i^\ast({\bf x})
\phi_j^\ast ({\bf y}) {\cal V}({\bf x}-{\bf y}) \phi_k({\bf y}) \phi_l({\bf x})
\end{equation}
which then specifies
\begin{eqnarray}
U&=& \langle ii| {\cal V} |ii \rangle \nonumber \\
V&=& \langle ij| {\cal V} |ji \rangle \nonumber \\
W&=& \langle ij| {\cal V} |ij \rangle \nonumber \\
X&=& \langle ii| {\cal V} |ij \rangle
\end{eqnarray}
where $i$ and $j$ are nearest neighbor sites.
The magnitudes of these terms have been discussed in the literature,
for example by Campbell, Gammel and Loh \cite{cgl}.
The qualitative conclusions are that while $U > V$, $W$, $X$, the relative
magnitudes
of the latter are sensitive to the detailed structure of the Wannier function
and of the effective (screened) interaction. To illustrate the origin
of this sensitivity we show in Fig.~1 a model calculation of the $\pi$-band
Wannier function for C$_{60}$. Generally $V>X>W$, but they
become comparable when the interaction is screened on length scales shorter
than the localization length of the Wannier function; indeed $X$ can even
change sign \cite{cgl}. We will rely upon these estimates when we return below
to the question of assigning values relevant to C$_{60}$.

\noindent
\section{Results for C$_{60}$ }

Exact diagonalization of our Hamiltonian for C$_{60}$ is
currently out of the question. Therefore, we have followed previous work
in computing the energies of the low-lying states (those which constitute
the degenerate ground state manifold in the absence of interactions) for
the anions C$_{60}^{-i}$ ($i=0-6$) perturbatively to second order in
the various interactions. Our procedure was to calculate {\em total\/}
energies for each ion. While this requires more computer time than
a more sophisticated approach (such as employed in \cite{cgk})
that calculates energy differences with respect to C$_{60}$ or
C$_{60}^{-6}$, it has the advantage of being straightforward to
code.  Comparing our results for the $U$ terms with previous work
gives a nontrivial check of the programs' validity.
We have not attempted to estimate directly
the neglected higher order corrections; instead we compare, below,
perturbation theory with exact diagonalization results for the smaller
homolog, C$_{12}$.

In Tables I--XIII we list the coefficient matrices for the $E_{i}$, $i=0-6$.
The degeneracies of the various anions can be understood, literally, in
a spherical approximation for C$_{60}$. The neutral molecule and C$_{60}^{-6}$
consist entirely of filled shells and have a unique ($L=0,S=0$) ground state.
The three degenerate lowest unoccupied molecular orbitals (LUMOs) of the
neutral C$_{60}$ molecule can be treated as an $L=1$ triplet. Consequently,
the low-lying states of C$_{60}^{-1}$ and C$_{60}^{-5}$ form an
($L=1$, $S=1/2$) multiplet, those of C$_{60}^{-2}$ and
C$_{60}^{-4}$ consist of the multiplets ($L=0$, $S=0$),
($L=1$, $S=1$) and ($L=2$, $S=0$) and finally
those of C$_{60}^{-3}$ consist of the multiplets ($L=0$, $S=3/2$),
($L=1$, $S=1/2$) and ($L=2$, $S=1/2$) \cite{cgk}. We note
that the $L=0$, 1, 2 states correspond respectively to the
icosahedral representations $^1A_g$, $^3T_{1g}$, $^1H_g$ for $n=2$ and 4,
and to $^4A_u$, $^2T_{1u}$, and $^2H_u$ for $n=3$.
Each entry in the tables is the coefficient of the term involving the
product of the its row and column labels in the expansion to second order.

Exploring the four-dimensional phase diagram implicit in these tables is a
formidable but unnecessary task, for the values of the
parameters are constrained by the requirement that perturbation theory
be valid and that they derive from a single underlying interaction. It
is convenient to consider families of interactions $(U,V,W,X)$ with
variable $U$ and fixed ratios $V/U$, $W/U$, $X/U$ where the first constraint
is incorporated (very roughly) by requiring that $U$ is bounded by the
$\pi$-orbital single-particle bandwidth ($\approx 5t$).
Based on the totality of the parameter values reviewed
in \cite{cgl} we estimate that the second constraint is incorporated
by considering $V/U \le 0.6$, $X/U \le 0.2$ and $W/U \le 0.04$.
At the
high end these values are consistent with estimates for benzene \cite{legend}
where the bare interaction is screened only by the $\sigma$ bands but in
that limit one would need to keep track of longer range interactions
\cite{fn4}. The problem of interest, however, is that of pair-binding in
the metallic phase of doped C$_{60}$, and ({\it modulo} self-consistency) we
take the interaction to be the effective, screened interaction appropriate
to this system, which will tend to give somewhat smaller values
of the parameters.

Before discussing our data, two remarks are in order. First, a caveat:
most of the estimates that we have cited in
assessing the relevant parameter ranges are from calculations using
atomic orbitals and not using Wannier functions. Campbell, Gammel and
Loh \cite{cgl} carried out a model calculation for a one dimensional
Kronig-Penney model where they calculated the Wannier function and found
systematic deviations from the atomic orbital estimates as the screening
length was varied. For our purposes the more interesting aspect of their
data is that they suggest that $W$ and $X$ might be substantially smaller
than we have supposed reasonable. Nevertheless, it is not at all obvious
whether this feature of their work, which is certainly sensitive to details
of the Wannier function and of the interaction, would carry over to a ``first
principles'' calculation for C$_{60}$. Hence we have chosen, pending a careful
estimate of the parameter values relevant to C$_{60}$, to use the atomic
orbital values as the appropriate ones. (Also, see our concluding
discussion.)
Second, the reader should note that particular results that do not
specify the
values of $V$, $W$ and $X$ will correspond to a ``canonical'' set
$V/U=0.5$, $W/U=0.04$ and $X/U=0.12$.

\subsection{Hund's Rules Violations}

We are interested here in Hund's first rule which implies that any
degeneracies arising from a partially filled shell in excess of those
dictated by symmetry be lifted in favor of the states with maximal spin.
For C$_{60}^{-n}$, $n=0$, 1, 5 and 6, the states are either nondegenerate
(0, 6) or degenerate by symmetry (1, 5). For the remaining anions we
find the following:

\noindent
(1) {\bf C$_{60}^{-2}$:}
At sufficiently small values of $U/t$, the ground state
is always the $^3T_{1g}$ state consistent with Hund's rule.
For the pure Hubbard model $V=W=X=0$, Hund's rule is violated
beyond $U_c = 2.8 t$ as the ground state crosses over to the $^1A_g$ state;
the $^1H_g$ state is always an excited state. (Here and elsewhere the
results for the pure Hubbard case are due to CGK and are listed for
comparison.) The inclusion of a nonzero $V$ does not shift the location
of the $^1A_g/^3T_{1u}$ level crossing very much and for $V/U \leq 0.5$
it remains at $U_c \approx 2.8 t$. The same is true of the off-diagonal terms,
including both $W$ and $X$ causes a modest downward shift in $U_c/t$ to about
2.7.

\noindent
(2) {\bf C$_{60}^{-3}$:} Here Hund's rule correctly predicts that the $^4A_u$
state is the ground state at small values of $U/t$. For the pure Hubbard
case there is a crossing to the $^3T_{1u}$ state at $U/t \approx 2.9$. We
find that this crossing is robust; even with the further inclusion of $V$,
$W$, and $X$ the crossing is always to the $^3T_{1u}$ state and the critical
value of $U/t$ is constant to within about 0.1.

\noindent
(3) {\bf C$_{60}^{-4}$:} This is, roughly, the particle-hole conjugate (within
the $t_{1u}$ manifold) of
C$_{60}^{-2}$ and the Hund's rule state at small $U/t$ is again the
$^3T_{1g}$ state. The pure
Hubbard result of a $^1A_g/^3T_{1g}$ level crossing at about $U/t \approx
2.8$ is mildly decreased by the addition of further interactions, with values
of $U_c$
= 2.5 (for $UV$), = 2.6 (for $UVW$), = 2.4 (for $UVX$), and = 2.6 (for $UVWX$).

\subsection{Pair-binding Energies}

The pair-binding energies for even dopings, {\em i.e.} for the
disproportionations
C$_{60}^{-2}$ $\rightarrow$ C$_{60}^{-1}$ + C$_{60}^{-3}$ and
C$_{60}^{-4}$ $\rightarrow$ C$_{60}^{-3}$ + C$_{60}^{-5}$, are always
negative and hence pair-binding does not occur in these cases. For the odd
dopings the situation is as follows.

\noindent
(1) {\bf $E_{\rm pair}^{(1)}$:} In the pure Hubbard model pair-binding takes
place only into the $^1A_g$ state of C$_{60}^{-2}$ and $E_{\rm pair}^{(1)}$
becomes positive for $U>U_{\rm pair} = 3.2 t$.
The value of $U_{\rm pair}$ is {\em extremely}
sensitive to the inclusion of further interactions. The inclusion
of the next neighbor repulsion at the values  $V/U=0.2$, 0.5 changes
$U_{\rm pair}/t$ to $4.2$ and $10$ respectively. Of course, finding
within perturbation theory a value of $U_{\rm pair}= 10t$ is
meaningless, except to suggest that $E_{\rm pair}$ is {\it never} positive.
This tendency of further neighbor density-density (``diagonal'')
interactions to suppress
pair-binding was the basis of Goff and Phillips's conclusion that going
beyond the pure Hubbard model was, for practical purposes, fatal to
the correlation mechanism for superconductivity.
For $E_{\rm pair}^{(1)}$, we find that the
inclusion of the off-diagonal terms, more precisely $W$, can substantially
affect $U_{\rm pair}$. For example, at $V/U=0.5$, the inclusion of
$W$ at the entirely
plausible level of $4\%$ of $U$ brings
$U_{\rm pair}$ down to $5.2t$, just outside
our nominal ``perturbative'' range. The further effect of an
added $X$ is mildly suppressive
of pair-binding, {\em e.g.} $X=0.12$ yields $U_{\rm pair} = 5.4 t$. These
results
are illustrated in Figs.~2, 3 and 4. Note that the ordering of the the
different
$E_{\rm pair}^{(1)}$ is also the ordering of the states of the C$_{60}^{-2}$.

\noindent
(2) {\bf $E_{\rm pair}^{(3)}$:} Here there are potentially several choices
for disproportionation, but in the pure Hubbard case and in the extensions
studied here, both the doubly and quadruply charged anions are in the
$^1A_g$ state {\em when} the pair-binding energy goes positive. The values
of $U_{\rm pair}/t$ for this channel show modest variation. It is suppressed
from the pure Hubbard value of 3.3 to the value 4.1 by the inclusion of
$V=0.5$.
Note that $V$ appears to be substantially less effective in
suppressing pair-binding here than in the case of $E_{\rm pair}^{(1)}$.
Again the inclusion of $W$ enhances the pairing and at $W/U=0.04$ changes
$U_{\rm pair}/t$ to about 3.6. The further inclusion of $X$ has a marginal
effect.

\noindent
(3) {\bf $E_{\rm pair}^{(5)}$:} For the pure Hubbard case $U_{\rm pair} = 3.3
t$
for the $^1A_g$ state (of C$_{60}^{-4}$)
and the value of $U_{\rm pair}$ for the $^1H_g$ state
lies slightly outside the physical range. We find that while additional
interactions greatly suppress pair-binding into the latter, their effect
on pair-binding into the $^1A_g$ state is minor. In particular, we find
the values $U_{\rm pair}/t = 4$, 3.9, 3.8, 3.7 on including $V$, $VX$, $VW$ and
$VWX$ respectively. This is in some contrast to the behavior of the (clearly
approximately) particle-hole conjugate quantity $E_{\rm pair}^{(1)}$.

\section{Results for C$_{12}$}
To assess the validity of second
order perturbation theory, particularly with regard to the sign of the
$E_{\rm pair}$, we have followed White {\it et al.}~\cite{white} and
compared
perturbative and exact results for C$_{12}$. The
latter has many features in common with C$_{60}$ \cite{white}, most notably a
degenerate triplet of LUMOs.
(Lammert and Rokhsar \cite{lamrok} have argued that the reasonable success of
second-order perturbation theory for C$_{12}$, in the sense of agreement with
exact diagonalization at modest values of $U$, may not carry over
to C$_{60}$.  While their argument may well be correct, we believe
it may not be relevant to the issue that is addressed below.)
Tables XIV--XVII list the coefficient matrices for $E_{\rm pair}^{(1)}$
in C$_{12}$.

White {\it et al.\ }showed that for the pure Hubbard model for C$_{12}$, while
perturbation theory reliably predicted $U_{\rm pair}$ (and even more reliably
predicted the value of $U$ for level-crossings) the typical value of $E_{\rm
pair}$
was considerably overestimated for $U>U_{\rm pair}$.
We find that the {\it additional\/} changes in the
$E_{\rm pair}$ due to $V$, $W$ and $X$ are fairly well described by the
perturbative results even when perturbation theory overestimates
$E_{\rm pair}(U;V=W=X=0)$ by a factor of three.
This is illustrated in Fig.~5 for the case of an added
$W$ term.  It should be emphasized that this result is not entirely
expected:  the additional change has three terms, one coming from
first-order perturbation theory, $-2W/3$, and two from second-order
in perturbation theory, $-UW/8$ and $0.2338W^2$.  The last term is
entirely negligible for the range of parameters in the plot.
For $U = 2t$, as in Fig.~5, the second term is 3/8 of the first, but
since $U$ is already large
enough that the $O(U^2)$ term alone substantially
overestimates $E_{\rm pair}$ (and so there are important contributions
from $O(U^3)$ and higher), it is not obvious that $O(U^2W)$ and higher terms
can be neglected in a calculation of the changes in $E_{\rm pair}$ due to
the addition of a $W$ term.

\section{Conclusions}

Our results show that the critical values
of $U/t$ for the Hund's rules violating level crossings in the doubly, triply
and quadruply charged anions are quite insensitive to the inclusion of
all the nearest-neighbor interactions. This suggests that the effects of
electron correlations might be significant even for isolated anions for
which pair-binding is certainly ruled out, since the energy differences
between different charge states are dominated by the ``molecular
capacitance'' energy.
In this connection the work of Negri {\it et al.}~\cite{Negri}
merits a close examination.  It is interesting they find
that the ground state of C$_{60}^{-2}$ is a $^1A_g$ state
rather than a $^3T_{1g}$ state. However, they also find that the next-lowest
state is of $^1A_u$ symmetry, which suggests that any useful perturbative
treatment may need to employ nearly-degenerate perturbation theory and allow
for occupancy of the $t_{1g}$ molecular orbitals (that lie roughly
1 eV above the $t_{1u}$ LUMOs) in the unperturbed states.

On the question of pair-binding we find that the $W$ term reduces
(using the crude measure of $U_{\rm pair}$) by about $50\%$ the suppression of
pair-binding produced by the $V$ term. We note that in contrast to
$U$ and $V$, the $W$ term favors pair-binding already at {\em first} order
and also favors pair-binding in second order, in each case with
large coefficients. However, these large {\it a priori} effects are
offset by the small numerical value of $W$ expected on physical
grounds in C$_{60}$. The
net effect of the $X$ term is weakly suppressive for the values of
interest \cite{fn41}.

Goff and Phillips had argued that going beyond the Hubbard model necessarily
suppressed pair-binding and that for parameter values relevant to the
fullerides $E_{\rm pair}$ was always negative. At a minimum we have shown
that there are interactions beyond the Hubbard approximation that do favor
pair-binding and, to use the framework of \cite{gp}, open a narrow window
of parameters for which this could actually take place in the doped fullerides.
However, we feel that such purely microscopic considerations ought not be
taken too seriously for two reasons. First, we believe that our calculations
illustrate that the issue of pair-binding is quite delicate and to draw
phase-diagrams with any confidence one would need to be certain that all
relevant interactions had been kept, their effects calculated accurately,
and the parameter values known to high precision. These are
daunting challenges for current theory.
Second, the problem of physical interest
involves {\it intra}molecular interactions that are necessarily renormalized by
{\it inter}molecular dynamics ({\em e.g.} screening) in a
self-consistent fashion \cite{fn5}.
Consequently, absent a solution of the full problem, it is difficult to
assign properly the relevant parameter values; {\em e.g.} a modest
enhancement of
$W$ could greatly enhance pair-binding. We are {\em not} arguing that
microscopics
can never settle these sorts of issues, merely that in this particular problem
the existence of a region of pair-binding in parameter space has been clearly
demonstrated in model calculations \cite{fn6} and that the additional
problem of locating precisely the parameters of the physical system does not
appear
amenable to first-principles solution. Consequently, it would appear that
consistency of the scenario with the totality of experiments is perhaps a
better approach.

\acknowledgements
We are grateful to S. Kivelson for useful discussions. This work was
supported in part by the NSF grant No. (DMR 89-20538),
an allocation of supercomputer time at the NCSA (DKC and HQL),
the MacArthur Chair at the University of Illinois, an NSF Young Investigator
Award
(MPG) and NSF grants Nos. DMR 91--22385 and
DMR 91--57018 (SLS).


\begin{figure}
\caption{
A model $p$ band $C$ Wannier function calculated by L\"owdin
orthogonalization keeping only a nearest neighbor overlap of $0.1$. The
amplitudes at the labeled sites are: 1) 1.324 \  2) -0.359 \  3) -0.384 \
4) 0.169 \ 5) 0.092 \  6) -0.123. Note the oscillations in the signs of the
amplitudes.
}
\label{fig1}
\end{figure}

\begin{figure}
\caption{
$E_{\rm pair}^{(1)}/t$ as a function of $U/t$ for the $^1A$ (circles),
$^3T$ (squares) and $^1H$ (diamonds) states for $V/U=W/U=X/U=0$.
}
\label{fig2}
\end{figure}

\begin{figure}
\caption{
$E_{\rm pair}^{(1)}/t$ as a function of $U/t$ for the $^1A$ (circles),
$^3T$ (squares) and $^1H$ (diamonds) states for $V/U=0.5$ and $W/U=X/U=0$.
}
\label{fig3}
\end{figure}

\begin{figure}
\caption{
$E_{\rm pair}^{(1)}/t$ as a function of $U/t$ for the $^1A$ (circles),
$^3T$ (squares) and $^1H$ (diamonds) states for $V/U=0.5$, $W/U=0.04$ and
$X/U=0.12$.
}
\label{fig4}
\end{figure}

\begin{figure}
\caption{
Effect of $W$ on pair binding for C$_{12}$ at $U=2$. The solid curve is the
perturbative result for $E_{\rm pair}^{(1)}/t$ for the $L=E, S=0$ state
as a function of $W$ for $U=2$ and the individual points are exact
diagonalization results.
}
\label{fig5}
\end{figure}



\begin{table}
\caption{C$_{60}$: $E_{0}$ coefficients.}
\begin{tabular}{ldddd} 
\ &      $U$    &      $V$     &    $W$     &  $X$       \\

1 & 15.000000   & 77.787661    & -16.725967 & 93.161593  \\

U & -0.787020 & 1.882569 & -0.058881 &  -0.089861   \\

V & \        & -4.603424 & 8.000289 & -0.041936  \\

W & \        & \         & -24.599166 & -0.156653  \\

X & \        & \         & \         & -1.766317\\
\end{tabular}
\end{table}


\begin{table}
\caption{C$_{60}$: $E_{1}$ coefficients.}
\begin{tabular}{ldddd} 
\ &      $U$    &      $V$     &    $W$     &  $X$       \\
1 &15.500000 &80.877467 &-20.264802  & 94.575722 \\

U &-0.778808 &1.859149 &-0.048072 &0.083824 \\

V & \           &-4.582528 &7.905386 &-0.139145 \\

W & \           & \            &-24.049537 &0.105234 \\

X & \           & \            & \          & -1.764327\\
\end{tabular}
\end{table}


\begin{table}
\caption{C$_{60}$: $E_{2}$ coefficients for the $^1A$ state.}
\begin{tabular}{ldddd} 
\ &      $U$    &      $V$     &    $W$     &  $X$       \\

1 & 16.05       & 84.022833    & -23.392517 & 95.962138 \\

U & -0.785960   & 1.831229     & -0.305086  & 0.272130  \\

V & \           &  -4.521729   &  7.889589  & -0.217773 \\

W & \           & \            & -24.433176 & 0.509186  \\

X & \           & \            & \          &-1.915823 \\
\end{tabular}
\end{table}

\begin{table}
\caption{C$_{60}$: $E_{2}$ coefficients for the $^3T$ state.}
\begin{tabular}{ldddd} 
\ &      $U$    &      $V$     &    $W$     &  $X$       \\
1 & 16.000000 & 84.014493 & -23.898078& 95.989851 \\

U & -0.768050 & 1.834397 & -0.038575 & 0.255951 \\

V & \           & -4.522097 &7.673967  &-0.235815 \\

W & \           & \            & -23.295537 &0.371069 \\

X & \           & \            & \          & -1.759702\\
\end{tabular}
\end{table}

\begin{table}
\caption{C$_{60}$: $E_{2}$ coefficients for the $^1H$ state.}
\begin{tabular}{ldddd} 
\ &      $U$    &      $V$     &    $W$     &  $X$       \\
1 & 16.020000& 84.017829& -23.695854& 95.978765 \\

U & -0.774057 & 1.834527 &-0.048615 & 0.260811 \\

V & \           & -4.521711& 7.674081& -0.230138\\

W & \           & \            &-23.490952 & 0.383309\\

X & \           & \            & \          &-1.797040 \\
\end{tabular}
\end{table}


\begin{table}
\caption{C$_{60}$: $E_{3}$ coefficients for the $^1A$ state.}
\begin{tabular}{ldddd} 
\ &      $U$    &      $V$     &    $W$     &  $X$       \\
1 & 16.500000 &87.198739 &-27.625793 &97.403980 \\

U &-0.754474 &1.808311 &-0.030389 &0.426520 \\

V & \           &-4.462598 &7.467910 &-0.331949 \\

W & \           & \            &-22.499046 &0.640850 \\

X & \           & \            & \          &-1.752442 \\
\end{tabular}
\end{table}

\begin{table}
\caption{C$_{60}$: $E_{3}$ coefficients for the $^2T$ state.}
\begin{tabular}{ldddd} 
\ &      $U$    &      $V$     &    $W$     &  $X$       \\
1 &16.550000 &87.207079 &-27.120232 &97.376267 \\

U &-0.771617 &1.806335 &-0.216324 &0.440273 \\

V & \           &-4.466601 &7.633590 &-0.315370 \\

W & \           & \            & -23.429825 &0.742408 \\

X & \           & \            & \          &-1.886450 \\
\end{tabular}
\end{table}

\begin{table}
\caption{C$_{60}$: $E_{3}$ coefficients for the $^1H$ state.}
\begin{tabular}{ldddd} 
\ &      $U$    &      $V$     &    $W$     &  $X$       \\
1 &16.530000 & 87.203743 & -27.322456 &97.387352 \\

U &-0.763712 &1.808521 &-0.045385 &0.433159 \\

V & \           &-4.464760 &7.481182 &-0.323543 \\

W & \           & \            &-22.797873 &0.658778 \\

X & \           & \            & \          &-1.807736 \\
\end{tabular}
\end{table}


\begin{table}
\caption{C$_{60}$: $E_{4}$ coefficients for the $^1A$ state.}
\begin{tabular}{ldddd} 
\ &      $U$    &      $V$     &    $W$     &  $X$       \\
1 &17.100000 &90.446885 &-30.436827 &98.762683 \\

U &-0.772486 &1.776993 &-0.395175 &0.620869 \\

V & \           &-4.417673 &7.648234 &-0.394751 \\

W & \           & \            &-23.526622 &1.116251 \\

X & \           & \            & \          &-2.008186 \\
\end{tabular}
\end{table}

\begin{table}
\caption{C$_{60}$: $E_{4}$ coefficients for the $^3T$ state.}
\begin{tabular}{ldddd} 
\ &      $U$    &      $V$     &    $W$     &  $X$       \\
1 &17.050000 &90.438544 &-30.942387 &98.790396 \\

U &-0.754729 &1.780108 &-0.128874 &0.606857 \\

V & \           &-4.408902 &7.388939 &-0.412432 \\

W & \           & \            &-22.369969 &0.979576 \\

X & \           & \            & \          &-1.854441 \\
\end{tabular}
\end{table}

\begin{table}
\caption{C$_{60}$: $E_{4}$ coefficients for the $^1H$ state.}
\begin{tabular}{ldddd} 
\ &      $U$    &      $V$     &    $W$     &  $X$       \\
1 &17.070000 &90.441881 &-30.740163 &98.779311 \\

U &-0.760674 &1.780258 &-0.138830 &0.610849 \\

V & \           &-4.412171 &7.406521 &-0.406900 \\

W & \           & \            &-22.572990 &0.991240 \\

X & \           & \            & \          &-1.890829 \\
\end{tabular}
\end{table}


\begin{table}
\caption{C$_{60}$: $E_{5}$ coefficients.}
\begin{tabular}{ldddd} 
\ &      $U$    &      $V$     &    $W$     &  $X$       \\
1 &17.600000 &93.725570 &-34.353421 &100.176812 \\

U &-0.752013 &1.750624 &-0.228459 &0.783467 \\

V & \           &-4.358270 &7.350974 &-0.492741 \\

W & \           & \            &-22.189387 &1.320806 \\

X & \           & \            & \          &-1.951428 \\
\end{tabular}
\end{table}

\begin{table}
\caption{C$_{60}$: $E_{6}$ coefficients.}
\begin{tabular}{ldddd} 
\ &      $U$    &      $V$     &    $W$     &  $X$       \\
1 &18.150000 &97.059816 & -37.858900 & 101.563228 \\

U &-0.746599 &1.719861 &-0.329144 &0.956353 \\

V & \           &-4.307167 &7.339871 &-0.572873 \\

W & \           & \            &-21.943153 &1.664541 \\

X & \           & \            & \          &-2.043404 \\
\end{tabular}
\end{table}


\begin{table}
\caption{C$_{12}$ $E_{\rm pair}^{(1)}$ coefficients for the $^1E$ state.}
\begin{tabular}{ldddd}
\ &      $U$   &      $V$    &    $W$    &  $X$       \\
1 & 0.0000 & $-0$.3333 & $-0$.6667 & 0.0000 \\
U & 0.0156 & 0.0452 & $-0$.1250 & $-0$.0764 \\
V & \  &  0.0428 & 0.2338 & $-0$.2535 \\
W & \  & \  & 0.2060 & 0.3056 \\
X & \  & \  & \  & 0.1320 \\
\end{tabular}
\end{table}

\begin{table}
\caption{C$_{12}$ $E_{\rm pair}^{(1)}$ coefficients for the $^3T_2$ state.}
\begin{tabular}{ldddd}
\ &      $U$   &      $V$    &    $W$    &  $X$       \\
1 & 0.0000 & $-0$.3333 & 0.6667 & 0.0000 \\
U & $-0$.0156 & 0.0313 & $-0$.1250 & 0.0764 \\
V & \  &  $-0$.0405 & 0.1273 & $-0$.0764 \\
W & \  & \  & $-0$.4329 & 0.2222 \\
X & \  & \  & \  & $-0$.0023 \\
\end{tabular}
\end{table}

\begin{table}
\caption{C$_{12}$ $E_{\rm pair}^{(1)}$ coefficients for the $^1A_1$ state.}
\begin{tabular}{ldddd}
\ &      $U$   &      $V$    &    $W$    &  $X$       \\
1 & $-0$.2500 & $-0$.0833 & $-1$.6667 & 1.0000 \\
U & 0.0339 & $-0$.0667 & $-1$.2482 & $-0$.2430 \\
V & \  &  0.0975 & $-0$.6273 & 0.0764 \\
W & \  & \  & 3.4074 & $-3$.5834 \\
X & \  & \  & \  & 1.0278 \\
\end{tabular}
\end{table}

\begin{table}
\caption{C$_{12}$ $E_{\rm pair}^{(1)}$ coefficients for the $^1T_1$ state.}
\begin{tabular}{ldddd}
\ &      $U$   &      $V$    &    $W$    &  $X$       \\
1 & $-0$.1667 & $-0$.1667 & 0.0000 & 0.6667 \\
U & $-0$.0041 & $-0$.0498 & 0.0697 & 0.0394 \\
V & \  &  0.0428 & 0.1042 & 0.0949 \\
W & \  & \  & 0.1042 & $-0$.4167 \\
X & \  & \  & \  & $-0$.0023 \\
\end{tabular}
\end{table}

\begin{references}
\bibitem{fn1} Actually their work (and ours) is intended to apply to a
molecule assumed to be embedded, self-consistently, in a metallic background.
\bibitem{fn2} We note that the data reported in \cite{cgk} are for a kinetic
energy ratio of 1.2 and thus are slightly different from those in this paper
for
the pure Hubbard model.
\bibitem{fn4} In that case we expect the longer ranged diagonal interactions
to dominate \cite{legend} and hence to find an attenuated version of the
pair-binding suppression found by Goff and Phillips.
\bibitem{fn41} For illustration we note that, $E_{\rm pair}^{(1)}$
($^1A_g$) = $
  -0.05\,U + 0.015364\,{U^2} - 0.0555602\,V + 0.0045\,U\,V -
   0.039903\,{V^2} - 0.41112\,W + 0.267823\,U\,W - 0.079106\,V\,W +
   0.933268\,{W^2} + 0.0277128\,X - 0.014621\,U\,X - 0.018581\,V\,X -
   0.142065\,W\,X + 0.153486\,{X^2}
$
\bibitem{fn5} In this context we note that there is an additional
complication, the dynamics of the screening, which has been discussed
recently by Lammert {\em et al} (SISSA listing cond-mat/9407071).
They conclude that this works in favor of pair binding.
\bibitem{fn6} All of this does assume that, as with smaller molecules,
second order perturbation theory for C$_{60}$ is not qualitatively unreliable.
\end{references}
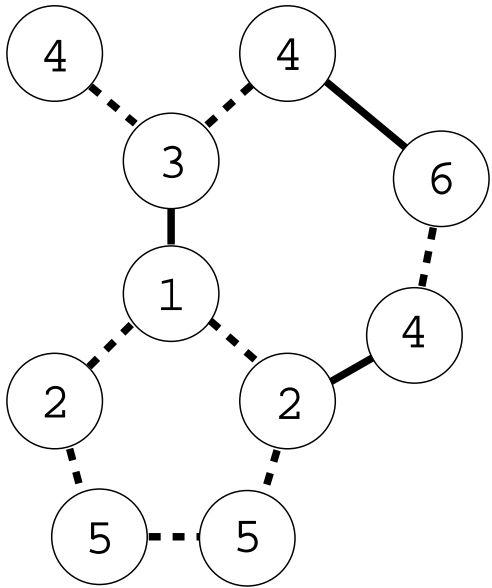
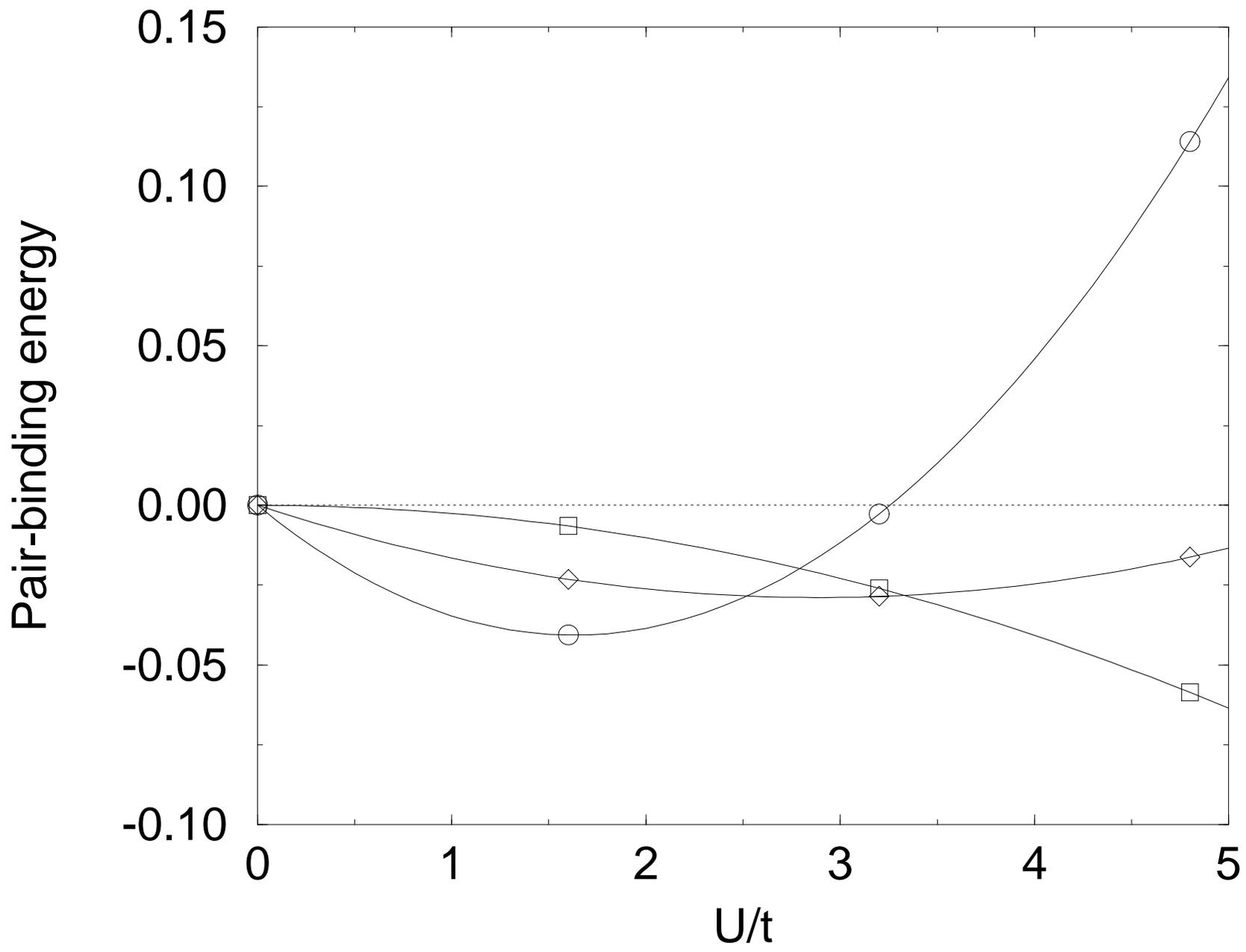
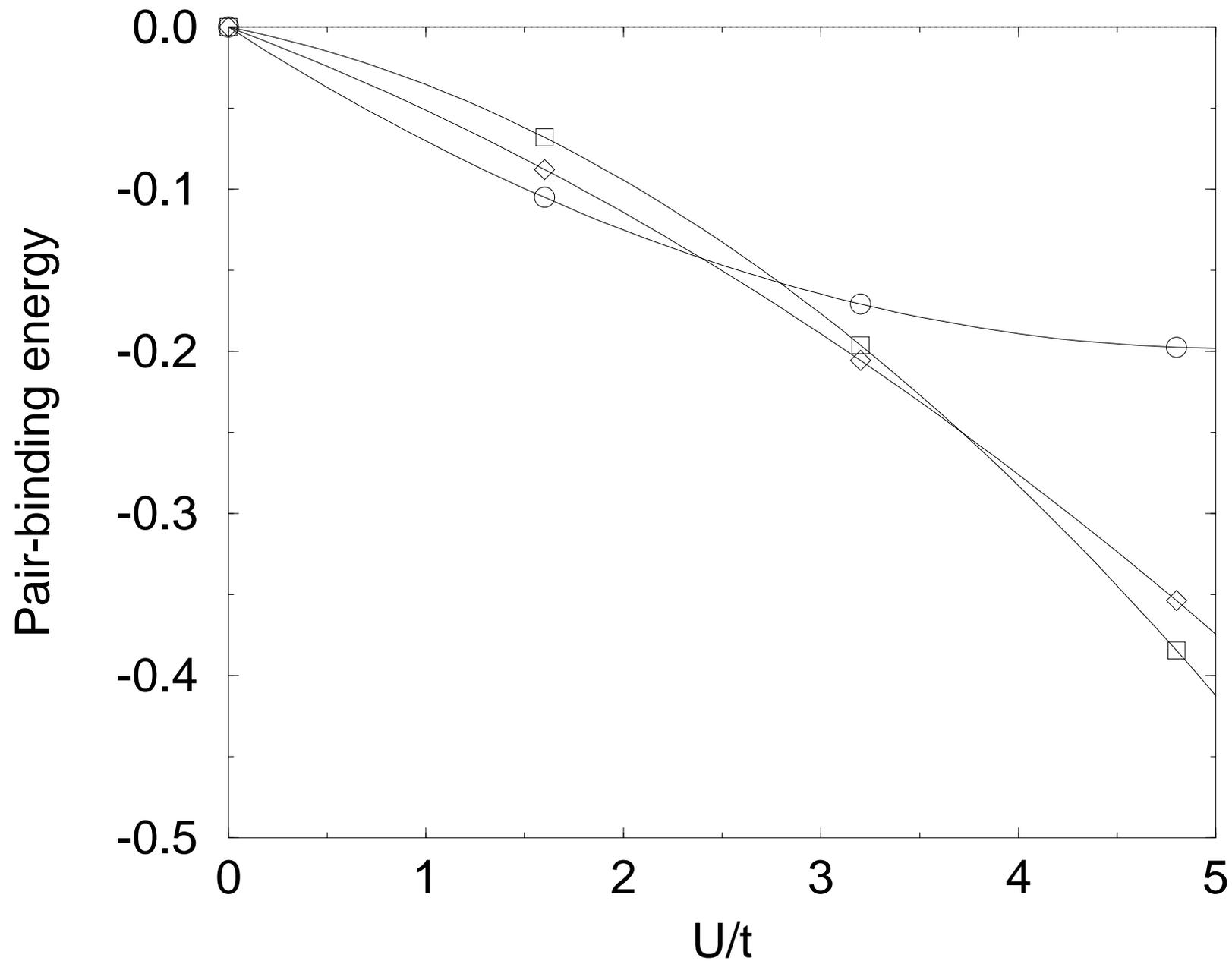
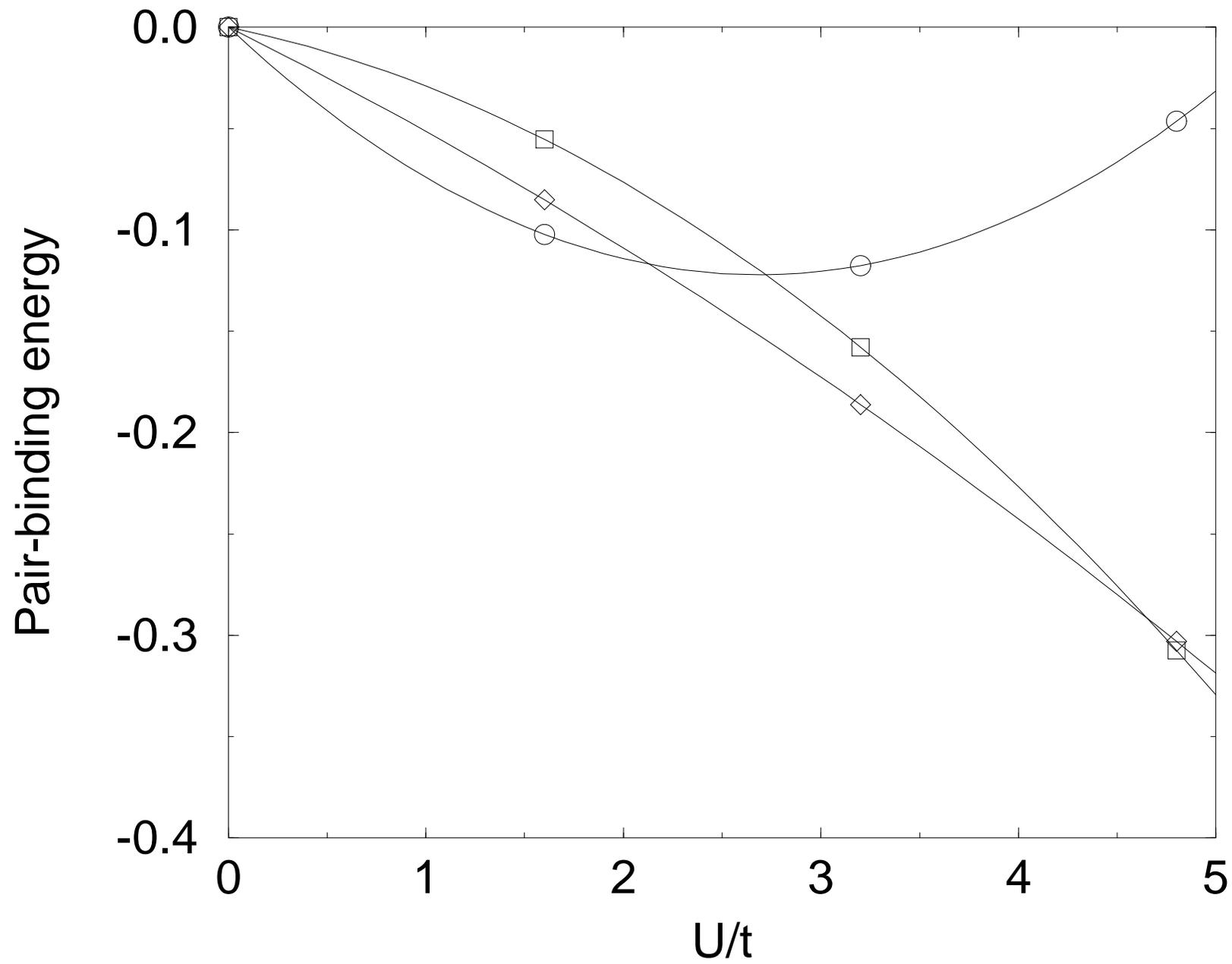
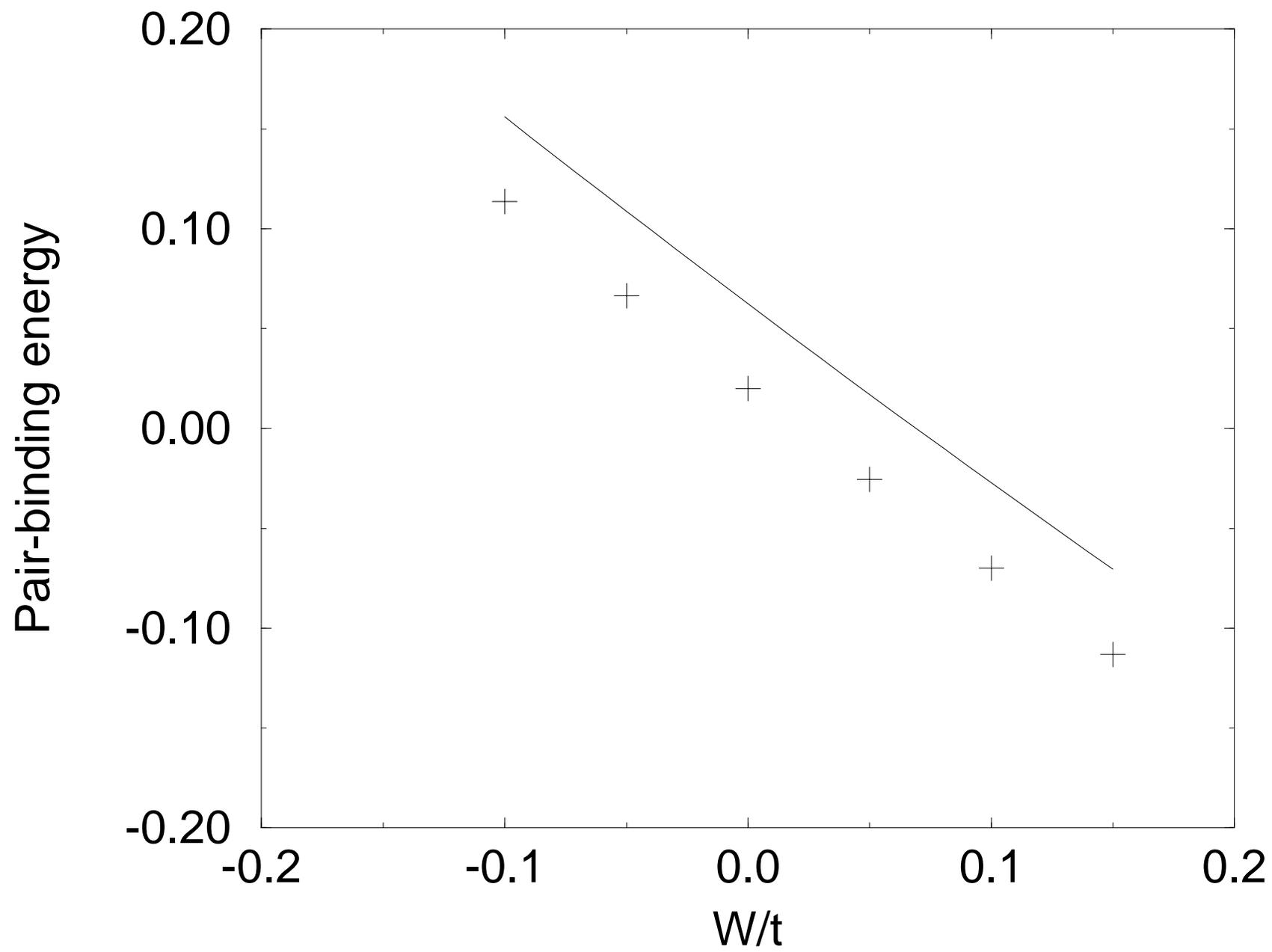
\end{document}